\documentclass{article}
\usepackage{setspace,graphicx,lscape,amsbsy,amsmath,amssymb,natbib}
\usepackage{epsfig}
\usepackage{mathtools}
\usepackage{amsthm}
\usepackage{bm}
\usepackage{algorithm}
\usepackage{algpseudocode}
\usepackage{graphicx}
\usepackage{cellspace, tabularx, booktabs}
\usepackage{hyperref}
\usepackage[english]{babel}
\usepackage[table]{xcolor}
\usepackage{graphicx}
\usepackage[utf8]{inputenc}
\usepackage[english]{babel}
\usepackage{alphabeta}
\usepackage[hmarginratio=1:1]{geometry}
\usepackage{booktabs}   
\usepackage{ltablex}
\usepackage[normalem]{ulem}

\hypersetup{
    colorlinks=true,
    linkcolor=blue,
    citecolor=blue
    }

\title{\textbf{Multinomial mixture for spatial data}}
\author{Anna Nalpantidi, Dimitris Karlis and Panagiotis Papastamoulis \\
Department of Statistics\\
Athens University of Economics and Business}
\begin{document}
\maketitle

\begin{abstract}
The purpose of this paper is to extend standard finite mixture models in the context of multinomial mixtures for spatial data, in order to cluster geographical units according to demographic characteristics. The spatial information is incorporated on the model through the mixing probabilities of each component. To be more specific, a Gibbs distribution is assumed for prior probabilities. In this way, assignment of each observation is  affected by neighbors' cluster and spatial dependence is included in the model. Estimation is based on a modified EM algorithm which is enriched by an extra, initial step for approximating the field. The simulated field algorithm is used in this initial step. The presented model will be used for clustering municipalities of Attica with respect to age distribution of residents. 
\end{abstract}


\section{Introduction}
Clustering regions according to demographic traits is a convenient way to summarize all the available information for an area and reveal hidden subgroups of regions that share common characteristics. More specifically, demographic traits and especially age and sex \citep{Poston2010} are the most important elements for describing a population. Consequently, dividing an area into groups according to demographics can provide a better insight for the population structure of the area, rather than examining each region individually. However, a plausible assumption is that adjacent areas are affected by common social and environmental conditions and as a result they tend to have similar demographic traits. This implies that data are spatially dependent taking into consideration the first law of geography \citep{Tobler1970}. As a consequence, standard clustering methods may not be adequate for analyzing these data, since they assume independent observations and ignore the extra variation due to spatial heterogeneity. 

Spatially constrained clustering is rapidly evolving in the last decades, as it is embedded in a wide range of fields like epidemiology, image processing criminology, geospatial economics and others. Constrained clustering was defined by \cite{Legendre1987} who was interested in including spatial, temporal or spatio-termporal restrictions. Since then, many popular clustering algorithms have been modified to adapt spatial information.  \cite{Pappas1992} presented a generalization of $k$-means in the context of image segmentation. He introduced a generalization of $k$-means, by using a Gibbs random field \citep{Geman1984, besag1986statistical,derin1987modeling} for inserting the spatial constraints. As hierarchical clustering is concerned, \cite{Carvalho2009} provided a modification of the classical hierarchical agglomerative method. More specifically, neighborhood restriction was imposed on the proximity matrix, which now refers to pairs of adjacent areas and not all possible pairs of observations. Due to the popularity of the field, an excessive number of algorithms have been proposed in the literature \citep{Liu2012,Ng2002,Kumar2012,Patil2006}.        

In the context of finite mixture models, \cite{Green2002} proposed a general approach for modelling spatial heterogeneity of count data instead of the model proposed by Besag, York, Molli{\`e} (known as the BYM model) \citep{Besag1991} and other similar models for the same purpose which use multivariate Gaussian distribution and incorporate spatial structure on covariance matrix. They proposed modelling allocation variables as Hidden Markov Random Fields (HMRF) that follows Potts model, with the purpose of facing local discontinuities and oversmoothed maps that may occur with the other methods.  

A few years later, \cite{Alfo2008}, based on the work of \cite{Green2002}, introduced an extension  of this model, in the context of image-segmentation. They proposed a Gaussian mixture model where allocation variables are modelled using a discrete HMRF. Then, due to the Hammersley-Clifford theorem, it is proved that component labels follow a Gibbs distribution. The difference with the previous work lies on the root of adopting a Gibbs distribution with cluster-specific parameters in prior probabilities.  While interaction parameter was constant for all the regions, in the work of \cite{Alfo2008} it is allowed to vary across the image/region. However, the concept of inhomogenous HMRF with spatially varying interaction parameter wasn't new since, \cite{Aykroyd1999} had proposed approaches on this framework. \cite{Forbes2013}, followed the idea of discrete HMRF model for disease mapping, but proposed a more complex structure on the interactions. On the work of \cite{Green2002} and \cite{Alfo2008} adjacent areas are penalized equally. \cite{Forbes2013} proposed taking into consideration, not only the adjacency of regions but also the adjacency of clusters.

In 2009, \cite{Alfo2009} extended their work \citep{Alfo2008} for mapping spatially dependent disease counts with special focus on multivariate disease counts. Based on these works, we propose a finite multinomial mixture model for spatial data. The presented model takes into account spatial dependence that leads to an extra source of variation, except for the common heterogeneity. In contrary to standard multinomial mixture model, this methodology let spatial dependence influence the component in which each observation will be assigned. The spatial information is adapted on prior probability of each component, which is let to be affected by the cluster of adjacent areas for each observation.

The paper consists of five sections. In \hyperref[sec2]{Section 2}, we will present standard multinomial mixture, a brief discussion upon its estimation method and the definition of number of clusters. In \hyperref[sec3]{Section 3} we will define the proposed model and we will focus on theoretical elements to support it. In addition, there are details provided about model selection procedure and the estimation, since common computational methods are not adequate for the proposed model.
Some simulation results to support the new model are provided in \hyperref[sec4]{Section 4}.
A real data application of the model on data from Greek census of 2001 is included in \hyperref[sec5]{Section 5}. Finally, \hyperref[sec6]{Section 6}  summarizes all remarkable conclusions of the survey.            

\section{Multinomial mixture model for independent observations}
\label{sec2}
It is assumed that a surrounding area can be partitioned in $K$ groups of sub-regions that share common demographic traits. More specifically, we would like to find clusters where area units have similar proportion of residents in each age group. For this purpose we will use finite mixture models and especially mixtures of multinomial distributions, which is a plausible assumption for this purpose. A brief description of the model is presented. 

Let $J$ be the number of categories (age groups) and $\bm{\lambda}=(\lambda_1,\lambda_2,...,\lambda_J)$ be the vector of probabilities of success for each category. It is obvious that $\sum_{j=1}^{J} \lambda_j=1$ and $0\le \lambda_j \le 1$. Let $\bm{y}=(y_1,\ldots,y_J)'$ be an observation from multinomial distribution. Then $m$ denotes the total number of records of $\bm{y}$: $m=\sum_{j=1}^{J} y_{j}$.  

Considering K clusters, there are $K$ multinomial distributions with corresponding probability vectors $\bm{\lambda_k}=(\lambda_k{}_1,\lambda_k{}_2,...,\lambda_k{}_J),~~k=1,\ldots,K$. Then  $\bm{y}$ is distributed according to the following probability distribution: 

\begin{equation}
P(\bm{Y})=\sum_{k=1}^{K} p_k \frac{m!}{y_1!y_2! \cdots y_J!}\lambda_{k1}^{y_{1}} \lambda_k{}_2^{y_{2}} \cdots \lambda_{kJ}^{y_{J}}=\sum_{k=1}^{K} p_k \frac{m!}{y_{1}!y_{2}! \cdots y_{J}!} \prod_{j=1}^{J}\lambda_{kj} ^{y_{j}}
\label{equation1}
\end{equation}

Taking into consideration the unobserved labels ($\bm{Z}$) that specify from which component each observation comes from we can write that
$$\bm{Z} \sim Multinomial(1,\bm{p})~~\mbox{and}$$
\begin{equation}
\label{mult_mix}
\bm{Y}|\bm{Z=z} \sim Multinomial(m,\bm{\lambda}_z)
\end{equation}

A necessary and sufficient condition for the generic identifiability of finite mixtures of multinomial distributions is the restriction $m\geqslant 2K - 1$ \citep{teicher1963, doi:10.1080/01621459.1964.10482176, titterington1985statistical, grun2008identifiability}. The reader is referred to \cite{Papastamoulis2023} for estimation of finite mixtures of multinomial distributions under a maximum likelihood as well as a Bayesian point of view.

\section{Multinomial mixture for spatially dependent observations}
\label{sec3}
\subsection{Definition}
Finite mixture models is an appropriate tool for capturing the heterogeneity of a population since, they divide it into subgroups in a way that observations of the same cluster are homogeneous, while observations from different clusters are different. However, the model of \hyperref[sec2]{Section 2} does not capture the spatial heterogeneity that exists when spatial data is the case. This means that there is an extra source of variation that is not taken into consideration and as a result the model is inadequate. Therefore, we have to incorporate spatial information in the model.

In the same concept, it is assumed that there is an area, which consists of $n$ sub-regions. For each sub-region, the number of residents in each age group is recorded and it is denoted by $\bm{y}_i=(y_{i1},y_{i2},...,y_{iJ})'$. Let $m_i$ denote the total counts of $i$-th region. In addition, an adjacency matrix (${\bf A}_{nxn}$) which reflects the spatial dependence, is defined as follows:
\[A_{ij}=
\left\{
	\begin{array}{ll}
	1,  & \mbox{if $i$  and $j$ are adjacent }  \\
    0,  & \mbox{else,} 
		\end{array}
\right.\]
where, $A_{ij}$ denotes the element of $i$-th row and $j$-th column of matrix ${\bf A}$. The adjacency can be considered in different ways, like using the distance between two area units. However, we consider that two regions are adjacent if they are spatially contiguous. Let $N_i$, $i=1,\ldots,n$ denote the number of neighbors of $i$-th region or else the neighborhood system. In order to assume $N_i$ as a neighbor system of $i$-th region it must be valid that:
\begin{enumerate}
    \item $i \notin N_i, \forall i \in N$ 
    \item $j \in N_j \leftrightarrow i \in N_j$
\end{enumerate}
Moreover, no further features have been recorded for each region.
 A general approach for modelling spatial dependence in FMMs was introduced by \cite{Green2002} and extended by \cite{Alfo2008}. This approach will be adopted in this work  but in the framework of multinomial mixture. 

 \cite{Geman1984}  proposed modelling the class process $z_{ik}$, $i=1,\ldots,n$, $k=1,\ldots,K$ by a Markov Random Field (MRF). Recalling the Hammersley-Clifford theorem, modelling indicator variables as a MRF is equivalent of using a Gibbs distribution for prior probabilities.
 Based on the \textit{Hammersley-Clifford theorem} the process $\{\bm{Z_i}: i \in N\}$ is a MRF, if and only if its joint distribution is a Gibbs distribution  \citep{Besag1974}.



Then the joint distribution of the MRF is given by
\[
P_G(\bm{z})  =W^{-1}\exp\{-\sum_{c}^{}V_c(\bm{z_c})\},
\]
where $W  = \sum_{z}\exp\{-\sum_{c}^{}V_c(\bm{z_c})\} $ is the normalizing constant or partitioning function, a sum over all possible values of $\bm{z}$. The sum on the $V(\cdot)$function (also known as the potential function) is over the set of cliques. The term clique refers to the set of nodes in which every pair is connected by an edge if MRF is seen as a graph. In simple words, a clique is a pair of regions that are spatially contiguous.  

In that case, the common prior  probabilities of the mixture model $p_k=P(z_{ik}=1)$ are replaced by the conditional prior probabilities: $t_{ik}=P(z_{ik}=1| \bm{z_j}, j \in N_i ).$ 
Given the assigned neighborhood system, the Gibbs distribution is defined by:
\begin{equation}
\label{eq:23}
t_{ij}=P(z_{ik}=1 | \bm{z_j}, j \in N_i ) \propto \exp\{\alpha_k-\beta_k\sum_{j \in N_i}^{}V(z_{ik} z_{jk}) \},
\end{equation}
for $i=1,\ldots,n, ~~k=1,\ldots,K$.Then (\ref{equation1}) is reformed as:
\begin{equation}
\label{eq:24}
P(\bm{Y_i})=\sum_{k=1}^{K} t_i{}_k \frac{m_i!}{y_{i1}!,y_{i2}! \cdots y_{iJ}!}\lambda_{k1}^{y_{i1}} \lambda_{k2}^{y_{i2}} \cdots \lambda_{kJ}^{y_{iJ}}=\sum_{k=1}^{K} t_{ik} \frac{m_i!}{y_{i1}!,y_{i2}! \cdots y_{iJ}!} \prod_{j=1}^{J}\lambda_{kj} ^{y_{ij}}
\end{equation}

Equation (\ref{eq:23}) is a Gibbs distribution with cluster specific parameters. It is supported that the standard model of \cite{Green2002}, where intercept ($\alpha$) and interaction term ($\beta$) are constant to the whole area, is restrictive and in many physical systems this assumption is violated, thus the cluster-specific Gibbs is preferred. We follow the suggested potential function of \cite{Alfo2008}, namely
\begin{equation}
\label{eq:25}
V(z_{ik} z_{jk})=
\left\{
	\begin{array}{ll}
	-1,  & \mbox{if  } z_{ik}z_{jk}=1  \\
    +1,  & \mbox{otherwise} 
		\end{array}
\right.
\end{equation}
This structure implies that neighbors of $i$-th region that belongs to other cluster than $k$ are penalized. Then prior probability of $k$-th component for $i$-th observation is increased when its neighbors belong there. 

Adopting the above potential function and posing  first cluster as the reference category, the rior conditional probabilities in (\ref{eq:23})  are equivalent to  
\begin{equation}
    \label{eq:26}
t_{ik} \propto \exp[\alpha_k+\beta_k(n_{ik}-n_{ik}^{c})],~~~~i=1,\ldots,n, ~~k=1,\ldots,K , 
\end{equation}
with $\alpha_1=\beta_1=0$ for ensuring identifiability. 
Parameter $\alpha_k,$ $k=1,\ldots,K$ accounts for the prior probability of $k$-th cluster without taking into consideration the neighbors. On the other hand, parameter $\beta_k$, $k=1,\ldots,K$ expresses how the cluster assignment of each observations is affected by the cluster of the  neighbors. This means that in case of $\bm{\beta}=(0,...,0)'$, then standard multinomial mixture given in (\ref{equation1}) occurs.  $n_{ik}$ and $n_{ik}^{c}$  denote the number of neighbors of $i$-th region that belong and not belong to $k$-th cluster respectively.
The expression (\ref{eq:26}) corresponds to a multinomial logit model for the prior conditional probabilities of cluster membership; it is also called as Strauss automodel \citep{strauss1977}.

In this point, it is important to mention  that prior probabilities $p_k$, $k=1,\ldots,K$, which have very natural and useful interpretation, are  difficult to be calculated under this approach.  

\subsection{Parameter Estimation and Model selection}
Selecting maximum likelihood approach for the estimation of parameters, due to the unobserved labels, EM algorithm is the appropriate tool. However, spatial dependence increase the complexity of the model and as a consequence it is necessary to apply approximations for making the algorithm tractable. For this reason, instead of defining the common likelihood, an approximation has been proposed by \cite{Besag1975} which is a pseudolikelihood. Pseudolikelihood allows defining the joint distribution of the MRF  by taking the product of conditional probabilities of observing $z_i$, given the values of all the other regions ($z_j, j \neq$ i). In this way, the calculation of  normalizing constant of joint Gibbs distribution,which demands all possible realizations of MRF can be avoided. Then, assuming also that $Y_i|Z_i$ are independent for $i=1,\ldots,n$,
the observed likelihood is approximated by:
\begin{equation}
    L_{obs}(\bm{\theta}|\bm{y})=\prod_{i=1}^{n} \sum_{k=1}^{K} t_{ik} \frac{m_i!}{y_{i1}!y_{i2}! \cdots y_{iJ}!} \lambda_{k1}^{y_{i1}} \cdots \lambda_{kJ}^{y_{iJ}}
\end{equation}
and the complete likelihood and log-likelihood are approximated by:
\begin{equation}
    L_c(\bm{\theta} | \bm{y},\bm{z}) = \prod_{i=1}^{n} \sum_{k=1}^{K} \left [t_{ik} \frac{m_i!}{y_{i1}!y_{i2}! \cdots y_{iJ}!} \lambda_{k1}^{y_{i1}} \cdots \lambda_{kJ}^{y_{iJ}}\right]^{z_{ik}}
\end{equation}
and 
\begin{equation}
    \begin{split}
l_c(\bm{\theta} | \bm{y},\bm{z}) = \sum_{i=1}^{n} \sum_{k=1}^{K} \left[  z_{ik} \log(t_{ik})+ z_{ik} \log \left(\frac{m_i!}{y_{i1}!y_{i2}! \ldots y_{iJ}!} \lambda_{k1}^{y_{i1}} \ldots \lambda_{kJ}^{y_{iJ}}\right) \right]=\\
=\sum_{i=1}^{n} \sum_{k=1}^{K} \left[ z_{ik} \log(t_{ik})+ z_{ik} \sum_{j=1}^{J}y_{ij}\log(\lambda_{kj}) \right]+constant
\end{split}
\end{equation}
The above form of likelihoods includes the conditional distributions of $\bm{z_i} | \bm{z_j}, i \neq j$ instead of the marginal distribution $\bm{z_i}$. In case of complete independence of $\bm{z_i}$, pseudolikelihoods coincide with true likelihoods.
Having approximated the quantities we need, EM algorithm can be applied as common:
\begin{itemize}
    \item [$-$] \textbf{\textit{E-step}}: Taking the expectation of approximated log-likelihood with respect to unobserved data $\bm{z_i}$, the approximated posterior probability of $i$-th observation belongs to $k$-th cluster is estimated by:
\begin{equation}
\hat{z_{ik}}=w_i{}_k=E[z_{ik}|\bm{\theta},\bm{y}]=\frac{t_{ik}f(\bm{y_i} |\bm{\lambda_k})}{\sum_{\ell=1}^{K}t_{i\ell} f(\bm{y_i} | \bm{\lambda}_\ell)}
\end{equation}
where $\bm{\lambda_k}=(\lambda_{k1},...,\lambda_{kJ})'$ and $f(\bm{y_i} | \bm{\lambda}_k)$ the probability mass function of multinomial distribution for $i$-th observation and $k$-th component. Now, the approximated expected complete log-likelihood is calculated by:
\begin{equation}
\label{equation.4}
Q=\sum_{i=1}^{n} \sum_{k=1}^{K} \left[ w_{ik} \log(t_{ik})+ w_{i{}_k} \sum_{j=1}^{J}y_{ij}\log(\lambda_{kj})  \right]
\end{equation}
    \item [$-$] \textbf{\textit{M-step}}: 
In this step approximated maximum likelihood estimators are provided, by maximizing (\ref{equation.4}) with respect to $\bm{\theta}=(\bm{\theta}_1,\ldots,\bm{\theta}_K)$, where 
$\bm{\theta}_k=({\bm{\theta}_{k,mult}},{\bm{\theta}_{k,Gibbs}})
$ and $
{\bm{\theta}_{k,Gibbs}} = (\alpha_k,\beta_k)
$ and ${\bm{\theta}_{k,mult}} = (\lambda_{k1},\ldots,\lambda_{kJ})$ for $k=1,\ldots,K$

\textbf{\textit{Parameters of Gibbs distribution:}}
We have to solve the equation system
\begin{equation}
\frac{\partial Q}{\partial (\alpha_k, \beta_k)}=\frac{\partial }{\partial (\alpha_k, \beta_k)}\sum_{i=1}^{n} \sum_{k=1}^{K} w_{ik} \log(t_{ik})=0,~~ k=2,\ldots,K
\end{equation}
However, this equation is extremely complex and numerical methods are suggested for its solution.\par
\textbf{\textit{Parameters of multinomial distributions:}}   
The maximization of Q under the constraint $\sum_{j=1}^{J} \lambda_{kj}=1$ is needed for finding the estimators of $\lambda_{kj}$. Thus,  the maximum pseudolikelihood estimates for parameters of each multinomial and every category are given by the formula: 
\begin{equation}
\label{eq:37}
\hat{\lambda_{kj}}=\frac{\sum_{i=1}^{n}w_{ik}y_{ij}}{\sum_{i=1}^{n}w_{ik}m_i} 
\end{equation}
for $k=1,\ldots,K , j=1,\ldots,J$
\end{itemize}

The parameter estimation in this model is not a trivial issue due to Gibbs distribution. In order to facilitate the procedure, pseudolikelihood was proposed. Then, conditional probabilities can be approximated by MCMC simulations. However, it has been proved that this approach is restrictive since it is inappropriate when strong dependence between regions is present. In addition, it is computationally demanding. For this reason, another approximation of complex complete likelihood was introduced by \cite{Qian1991} and extended by \cite{Celeux2003}. This, method uses Mean field theory, from the domain of statistical mechanics, in order to approximate the HMRF and then apply EM. Mean field approximation means that the value of $i$-th area is independent to values of other areas while these are set to be fixed an irrelevant to $i$-th value. Based on this idea, intractable distributions like Gibbs distribution can be faced. 

According to this, \cite{Celeux2003} suggested a class of algorithms that consists of two steps. In the first stage, the hidden field (\textbf{z}) is approximated given the observed data ($\bm{y}$) and the current parameters  ($\bm{\theta}^{(t)}$), thus an approximation of Gibbs distribution is achieved as follows:
Denoting $\tilde{\bm{z}}$ the configuration of HMRF, 
$$P_{\tilde{\bm{z}}}(\bm{z}|\bm{\theta}_{Gibbs})=\prod_{i=1}^{n}P_G(\bm{z_i}|\tilde{\bm{z_j}},j \in N_i,\bm{\theta}_{Gibbs}^{(t)})$$
In the second step, given the corresponding approximated field, the approximated posterior probabilities can be estimated (E-step) and then M-step follows for providing estimations of the parameters of interest. They proposed three algorithms for approximating the hidden field: mean field, mode field and simulated field algorithm. The concept of simulated field algorithm will be used in this work as it has been proved that it provides better results. This is the counterpart of the Stochastic EM algorithm in a spatial framework.

The modified classification EM (CEM) algorithm  that used for estimating the proposed model can be sketched as follows. 
At the $t$-th iteration we have:
\begin{enumerate}
    \item Simulate $\tilde{\bm{z_i}}$ from $P_G(\bm{z_i}|\tilde{\bm{z_j}}^{(t-1)},j \in N_i,\bm{\theta}_{Gibbs}^{(t)})$ for $i=1,\ldots,n$
    \item Using the field approximation $\tilde{\bm{z}}$  from the previous step apply CEM algorithm
    \begin{enumerate}
    \item \textit{E-step}: Estimate posterior probabilites:
\begin{equation}
\label{eq:39}    
\tilde{w}_{ik}^{(t)}=\frac{t_{ik}(\tilde{\bm{z_i}}^{(t)} | \bm{\alpha}^{(t-1)}, \bm{\beta}^{(t-1)})f(\bm{y_i}|\bm{\lambda_k}^{(t-1)})}{\sum\limits_{l=1}^{K}t_{il}(\tilde{\bm{z_i}}^{(t)} | \bm{\alpha}^{(t-1)}, \bm{\beta}^{(t-1)} ) f(\bm{y_i}| \bm{\lambda}_l^{(t-1)})}
\end{equation}\par
Then, approximated expected complete log-likelihood is modified as:
\begin{equation}
\label{eq:38}
Q^{MF}=\sum_{i=1}^{n} \sum_{k=1}^{K} \left[ \tilde{w}_{ik}^{(t)}\log(t_{ik})+ \tilde{w}_{ik}^{(t)}\sum_{j=1}^{J}y_{ij}\log(\lambda_{kj}^{(t-1)}) \right]
\end{equation} 
    \item \textit{C-step}: Set $z_{ik}^{(t)}=1$  and all other 0,  if  $\tilde{w}_{ik}^{(t)}=max_{\ell}\tilde{w}_{i\ell}^{(t)}$, $\ell=1,\ldots,K$
    \item \textit{M-step}: Estimate 
   $\alpha_k^{(t)}, \beta_k^{(t)}$ and $\lambda_{kj}^{(t)}$ given $z_{ik}^{(t)}$ and $\tilde{\bm{w_i}}^{(t)}$  , $k=1,\ldots,K$ and $j=1,\ldots,J$.
    \end{enumerate}
\end{enumerate}

The above procedure is repeated for a significant number of iterations, until convergence to maximum observed approximated likelihood is achieved. 
The approximated observed likelihood is:
\begin{equation}
\label{eq:40}
L_{obs,approx}(\bm{\theta})=\prod_{i=1}^{n}\sum_{k=1}^{K}f(y_i|z_{ik},\bm{\lambda}_{k})P(z_{ik}=1|\tilde{\bm{z_j}},j \in N_i, \alpha_k,\beta_k)
\end{equation}
Up to this approximation of observed likelihood, \cite{Forbes2003} displayed an approximation of BIC in case of mean field  approximations.  Denoting $\hat{\bm{\theta}}$ as the approximation of true MLE then, the approximation of BIC is:

\begin{equation}
\label{eq:41}
BIC^{\tilde{\bm{z}}}(\hat{\bm{\theta}} )= 2\log(L_{obs,approx}(\hat{\bm{\theta}}))-d\log(n)
\end{equation}
where $d$ is the number of free parameters in the model, in our case $d=2(K-1)+K(J-1)$. The approximation of observed likelihood is given by the equation (\ref{eq:40}).  

Initialization of EM is a crucial issue, since it is possible to find a local maximum instead of the global maximum which is the case. For this reason, we propose the strategy of \cite{Biernacki2003}. More specifically, we used different initial values and ran the EM algorithm. For each set of initial values the maximum likelihood was recorded. The solution that corresponds to the highest approximated observed likelihood was used as initial value for the final EM. Finally as far as the convergence of the algorithm and because of the stochastic nature of the algorithm, we stopped when we failed to find a larger log-likelihood for 50 successive iterations.

\section{Simulation study}
\label{sec4}
In this section we examine the performance of the proposed model under different conditions. More specifically, we focus on different lattice sizes and various parameters of the Gibbs distribution, which measures how interaction of neighbors affects the assignment of each observation in each cluster. 
We assumed that we have a $N$-dimensional square lattice  ($N=8,10,20$). The points of each lattice come from a mixture of two multinomial distributions. For each multinomial distribution we have assumed 10 categories, and size equal to 100 
while the parameters are presented in Table \ref{tab:Sim}. According to the presented model, the prior probabilities of the mixture follow the Gibbs distribution, so we have three different scenarios for this distribution namely ${\bm \beta}=(0,0.01)',(0,0.1)'$ and $(0,0.2)'$ respectively. 

\begin{table}
    \centering
    \begin{tabular}{ccccccccccc}
      \hline
&Cluster 1 & Cluster 2\\
\midrule
$\lambda_{k1}$ & 0.12 & 0.08\\
$\lambda_{k2}$ & 0.12 & 0.08\\
$\lambda_{k3}$ & 0.12 & 0.08\\
$\lambda_{k4}$ & 0.12 & 0.08\\
$\lambda_{k5}$ & 0.12 & 0.08\\
$\lambda_{k6}$ & 0.08 & 0.12\\
$\lambda_{k7}$ & 0.08 & 0.12\\
$\lambda_{k8}$ & 0.08 & 0.12\\
$\lambda_{k9}$ & 0.08 & 0.12\\
$\lambda_{k10}$ & 0.08 & 0.12\\
\bottomrule 
 \end{tabular}
    \caption{Parameters of the multinomial distributions}
    \label{tab:Sim}
\end{table}

To evaluate the performance of the model we used the Adjusted Rand Index (ARI) \citep{hubert1985comparing}. The index measures the agreement between two partitions. Values close to $1$ indicate complete agreement. In this case, we compared the cluster labels estimated by the proposed model with the true labels. True labels have been simulated by a Gibbs sampler.

\begin{figure}[ht]
\begin{center}
\includegraphics[scale=0.5]{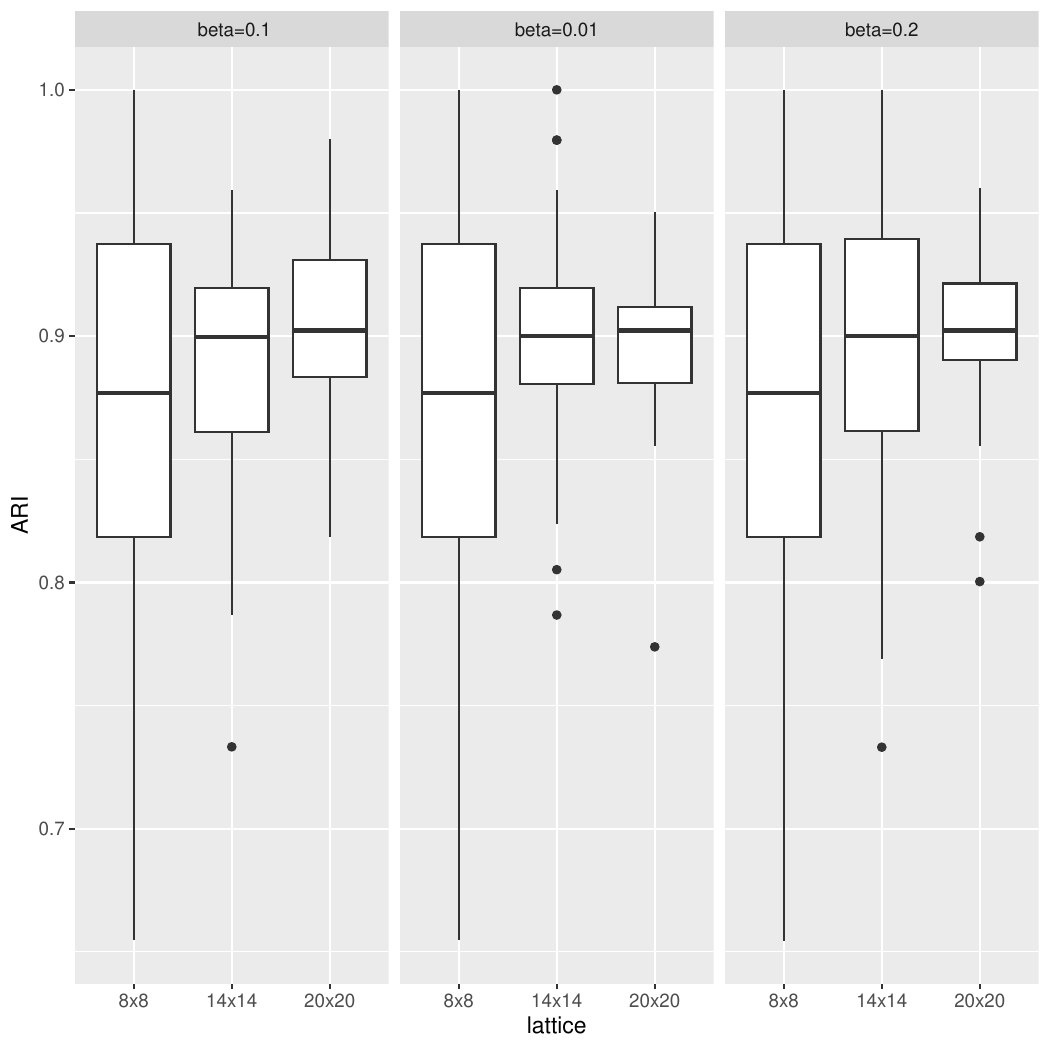}
\caption{Boxplots of ARI for different lattices and different values of $\beta$. \label{sim001}}
\end{center}
\end{figure}

For all scenarios we have simulated 100 samples. Figure \ref{sim001} depicts the results of the simulations. More specifically, we conclude that Gibbs distribution does not affect essentially the model's performance. To be more specific, ARI values are similar for each combination of $\beta$ parameter and sample size. On the other hand, comparing results with respect to different lattice size, it seems that there are significant differences: as sample size increases, the variability of the ARI decreases. When $N=8$, there are cases where ARI is less than $0.70$, while in case of $N=20$, ARI is never less than $0.80$. So, for small sample sizes it is possible not to have such a good partition as in higher sample sizes. We also see a slight improvement on ARI as lattice size increases. However, the general conclusion is that model clusters adequately the observations, since, in each case the median ARI is higher than $0.80$.

Finally note that initialisation of the algorithm as well as convergence were used as described in Section 3.

\section{Application: Data from Greek census of 2001}
\label{sec5}
Attica is one of the thirteen administrative regions of Greece which encompasses the capital of Greece, Athens, and the biggest port of the country, Piraeus. Except for Athens and surrounding cities, Saronic islands, Cythera and a small part of Peloponnese mainland pertain to Attica region. As it is expected, the extension of the region, in combination with the wave of urbanization, has led Attica to encompass a significant percentage of Greek population. To be more specific, according to Greek census of 2001, more than the one third of total Greek population lives in Athens, with the percentage estimated around 35\%. With the purpose of a better administration of Attica, it has been partition into 66 municipalities.

 It is common that people choose living in areas that service their needs and usually, their needs depend on their age. For example, young people prefer living in big cities instead of villages, in order to study or work. This, in combination with the fact that municipalities of Attica not only do they correspond to urban cities and small towns, but also islands and suburbs, motivates in believing that sub-regions of Attica present heterogeneity with respect to the age  distribution of their inhabitants. In other words, it seems plausible assuming that all municipalities of Attica do not present the same percentage of people in each age group, but some of them tend to be similar and described as a cluster. Furthermore, another hypothesis is that adjacent sub-regions are affected by similar environmental and social factors, so they attract people of the same age and consequently, this may encourage clustering on the same group. 
 
 The purpose of this section is to present a clustering of municipalities of Attica region according to age distribution of their residents. Use of similar data has been treated in \cite{jorgensen2013forming}
 where the spatial aspect was ignored. 
  Nevertheless, special interest is given to spatial dependence of the sub-regions of Attica. It is assumed that two regions are adjacent if they are spatially contiguous. For this reason, six area units that are far away from the main part of Attica (Cythera, Hydra, Spetses, Poros, Troezen, Agistri) have been excluded, since the main assumption of common characteristics due to adjacency is not plausible. The islands of Salamina and Aegina have been kept and as adjacency is concerned, it has been considered that Salamina is neighbor of Perama and Aegina is neighbor of Pireus, as long as the communication of the mainland of Attica with these islands becomes through the ports of the correspond municipalities.

 The clustering was conducted by applying the multinomial mixture with Gibbs prior, that was presented in section 3. The results were compared to the results of standard multinomial mixture. The analysis was conducted with \texttt{R} package.
 
The clustering refers to 60 of 66 municipalities of Attica. For each of them the number of inhabitants per age group is recorded. There are 18 age groups: from 0 to 84. Each group contains five ages and the final group refers to  people whose age is 85 and over. An example of the form of the data is presented in Table \ref{tab:1}.
The data comes from Greek census of 2001 and data were taken from the National Statistical Office website. 

\begin{table}
    \centering
    \begin{tabular}{ccc}
      \hline
Age group & Records & Proportions\\
\midrule
0-4 & 3132 & 0.038\\
5-9 & 3128 & 0.038\\
10-14 & 3387 & 0.042\\
15-19 & 6069 & 0.075\\
20-24 & 9304 & 0.114\\
25-29 & 7368 & 0.090\\
30-34 & 6770 & 0.083\\
35-39 & 5587 & 0.069\\
40-44 & 5595 & 0.069\\
45-49 & 5538 & 0.068\\
50-54 & 5244 & 0.064\\
55-59 & 4150 & 0.051\\
60-64 & 4400 & 0.054\\
65-69 & 3924 & 0.048\\
70-74 & 3429 & 0.042\\
75-79 & 2166 & 0.027\\
80-84 & 1253 & 0.015\\
85+ & 970 & 0.012\\
\bottomrule 
 \end{tabular}
    \caption{Number and rounded proportion of residents by age in Zografos from census of 2001 }
    \label{tab:1}
\end{table}

\begin{figure}[ht]
\begin{center}
\includegraphics[scale=0.42]{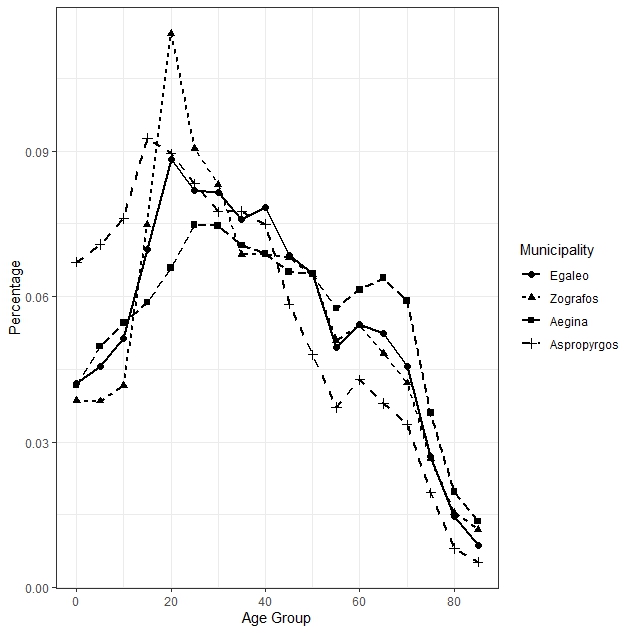}
\caption{Percentage of residents in each age group for four municipalities \label{pub1}}
\end{center}
\end{figure}

Before starting clustering, it would be interesting examining whether the hypothesis of heterogeneity and spatial dependence are plausible for this data. In Figure \ref{pub1}, a small number of municipalities have been chosen in order to indicate the heterogeneity of the regions. As it is obvious, not all regions have the same age profile. More specifically, Zografos presents an excess of records in ages 20-29. Furthermore, Aegina have high percentages of residents in ages 60+, compared to the other regions. On the other hand, Aspropyrgos dominate in younger ages of 0-19, while it presents the lower percentages on older age groups. Egaleo seems to represent an intermediate situation. 

After that, we examine the second assumption that regions are spatially correlated. For this purpose Moran's Index $I$ \citep{Moran1950} was used. The above index was used for examining spatial autocorrelation on univariate observations. However, in this study, observations are multivariate. For overcoming this problem, the mean age of each municipality is computed and used for the Moran index. Using mean age as variable of interest and the adjacency matrix as the weight matrix, the results shown that $I=0.33$ and there is evidence that municipalities of Attica are shown spatial correlation  (permutation test pvalue $<$ 0.001). 

The above study indicated that data presents heterogeneity so clustering is meaningful. First, standard multinomial mixture was applied for clustering data. The estimation of both models was based on the spatial counterpart of stochastic EM that was described in section 3. Especially for the case of standard multinomial mixture, the vector of spatial parameters ($\bm{b}_{k \times k}$) were set equal to zero. The two models are nested and this justify the usage of the same algorithm. Furthermore, the estimation of Gibbs parameters was based on simulated annealing algorithm. Variant of Newton-Raphson was also used, but in terms of maximum log-likelihood simulated annealing was slightly better. The algorithm applied for adequate number of iterations for each case, until convergence was achieved. Initialization process was based on the strategy described in Section 3 for $k=2$ while for $K=3,
\ldots,9$, initialization was based on the solutions of previous model plus a random vector for the new component. 
Convergence was detected as described at the end of Section 3. 

The maximum log-likelihood values as well as BIC for models with different number of components, from the application of standard multinomial mixture are provided in Table \ref{tab.2}. Maximum value of BIC indicates the optimal model. According to Table \ref{tab.2}, BIC suggests the model with eight components as the best choice. 

\begin{table}[ht]
    \centering
    \begin{tabular}{ccc}
    \toprule
\textbf{K} & \textbf{log-likelihood} & \textbf{BIC} \\
 \midrule
1 & -28590.73 & -57251.07 \\ 
2 & -18303.01 & -36749.32 \\
3 & -15971.25 & -32159.50\\
4 & -13096.92 &  -26484.54  \\
5 & -11878.35 & -24121.10  \\
6 & -11537.93 & -23513.95 \\
7 & -10730.52 & -21972.83\\
8 & -10670.59 &  -21926.79 \\
9 & -10658.97 &  -21977.13 \\
\bottomrule
    \end{tabular}
    \caption{ Maximized log-likelihood and BIC for  standard multinomial finite mixture \label{tab.2}}
\end{table}

The correspond results of the proposed model of multinomial mixture that takes into consideration the spatial information are presented in Tables \ref{tab.4} and \ref{tab.5}.

\begin{table}[ht]
    \centering
    \begin{tabular}{ccc}
    \toprule
\textbf{K} & $\textbf{log-likelihood}_{\textit{approx}}$ & $\textbf{BIC}_{\textit{approx}}$  \\
 \midrule
1 & -28590.73 & -57251.07  \\ 
2 & -18298.02 &  -36743.44 \\
3 & -15963.93 &  -32153.05 \\
4 & -13099.68 &  -26502.34\\
5 & -11873.5 &  -24127.77 \\
6 & -11525.79 &  -23510.15\\
7 & -10718.53 &  -21973.42 \\
8 & -10659.71 & -21933.57\\
9 & -10627.76 &  -21947.46\\
\bottomrule
    \end{tabular}
    \caption{Approximated maximized log-likelihood and BIC for  multinomial finite mixture with Gibbs prior \label{tab.4}}
\end{table}

\begin{table}[ht]
  \centering 
  \begin{tabular}{ccccccccc}
  \hline
 Parameter & 1st & 2nd & 3rd & 4th & 5th & 6th & 7th & 8th \\
 \midrule 
$\lambda_{k1}$ & 0.044 & 0.051 & 0.045 & 0.038 & 0.060 & 0.051 & 0.043 & 0.042\\
$\lambda_{k2}$ & 0.046 & 0.052 & 0.049 & 0.037 & 0.062 & 0.052 & 0.044 & 0.050 \\
$\lambda_{k3}$ & 0.050 & 0.056 & 0.051 & 0.040 & 0.066 & 0.055 & 0.045 & 0.055 \\
$\lambda_{k4}$ & 0.062 & 0.067 & 0.066 & 0.060 & 0.081 & 0.067 & 0.055 & 0.059 \\
$\lambda_{k5}$ & 0.076 & 0.073 & 0.079 & 0.090 & 0.088 & 0.080 & 0.065 & 0.066\\
$\lambda_{k6}$ & 0.080 & 0.074 & 0.080 & 0.092 & 0.089 & 0.085 & 0.068 & 0.075\\
$\lambda_{k7}$ & 0.083 & 0.083 & 0.079 & 0.088 & 0.085 & 0.091 & 0.076 & 0.075 \\
$\lambda_{k8}$ & 0.075 & 0.078 & 0.069 & 0.073 & 0.077 & 0.08 & 0.074 & 0.071\\
$\lambda_{k9}$ & 0.076 & 0.081 & 0.067 & 0.074 & 0.075 & 0.078 & 0.078 & 0.069\\
$\lambda_{k10}$ & 0.071 & 0.078 & 0.063 & 0.069 & 0.067 & 0.068 & 0.075 & 0.065 \\
$\lambda_{k11}$ & 0.071 & 0.075 & 0.068 & 0.067 & 0.059 & 0.064 & 0.076 & 0.065  \\
$\lambda_{k12}$ & 0.054 & 0.056 & 0.058 & 0.051 & 0.045 & 0.048 & 0.058 &  0.058\\
$\lambda_{k13}$ & 0.056 & 0.052 & 0.064 & 0.054 &  0.046 & 0.051 & 0.059 & 0.061 \\
$\lambda_{k14}$ & 0.052 & 0.043 & 0.060 & 0.051 & 0.038 & 0.048 & 0.055 & 0.064\\
$\lambda_{k15}$ & 0.047 & 0.035 & 0.049 & 0.048 & 0.030 & 0.039 & 0.053 & 0.059 \\
$\lambda_{k16}$ & 0.029 & 0.022 & 0.028 & 0.033 & 0.017 & 0.022 & 0.037 & 0.036 \\
$\lambda_{k17}$ & 0.016 & 0.013 & 0.015 & 0.020 & 0.009 & 0.012 & 0.022 & 0.020\\
$\lambda_{k18}$ & 0.012 & 0.010 & 0.010 & 0.015 & 0.006 & 0.009 & 0.016 & 0.014 \\
$\alpha_k$ &  0 & 2.048 &  1.572 &  0.600 &  1.074 &  0.938 &  0.408 & -0.642 \\
$\beta_k$ & 0 & 0.781 & 0.786 & 0.429 & 0.678 & 0.265 & 0.404 & 0.426\\
$n_k$  & 15 & 8 & 8 & 3 & 5 & 17 & 3 &  1\\
\bottomrule
 \end{tabular} 
  \caption{Parameter estimates for  multinomial finite mixture with Gibbs prior (rounded on 3 digits) \label{tab.5}}
\end{table}

It occurs that both models proposed the same number of clusters, while the estimated parameters of multinomial distributions are close. This is something expected, since this model is proposed for capturing the extra variation from spatial information that it is not explained by the standard model. In \cite{Celeux2003}, a simulation study indicates similar superior behavior of the spatial model, namely an improved misclassification rate, while the parameters estimates are similar.
They conclude that standard EM provides the same estimations of parameters with the modified EM that includes an extra step of field's simulation, but the second method presents lower error rate than the first. As it has been discussed, the two models are nested, so likelihood ratio test (LRT) can be used for comparing them. The results showed that the there is statistically significant difference between the two models which means that spatial term is statistically significant (LRT pvalue $<$ 0.05).

\begin{figure}[htbp]
\hspace*{-1.5cm}
\includegraphics[scale=0.4]{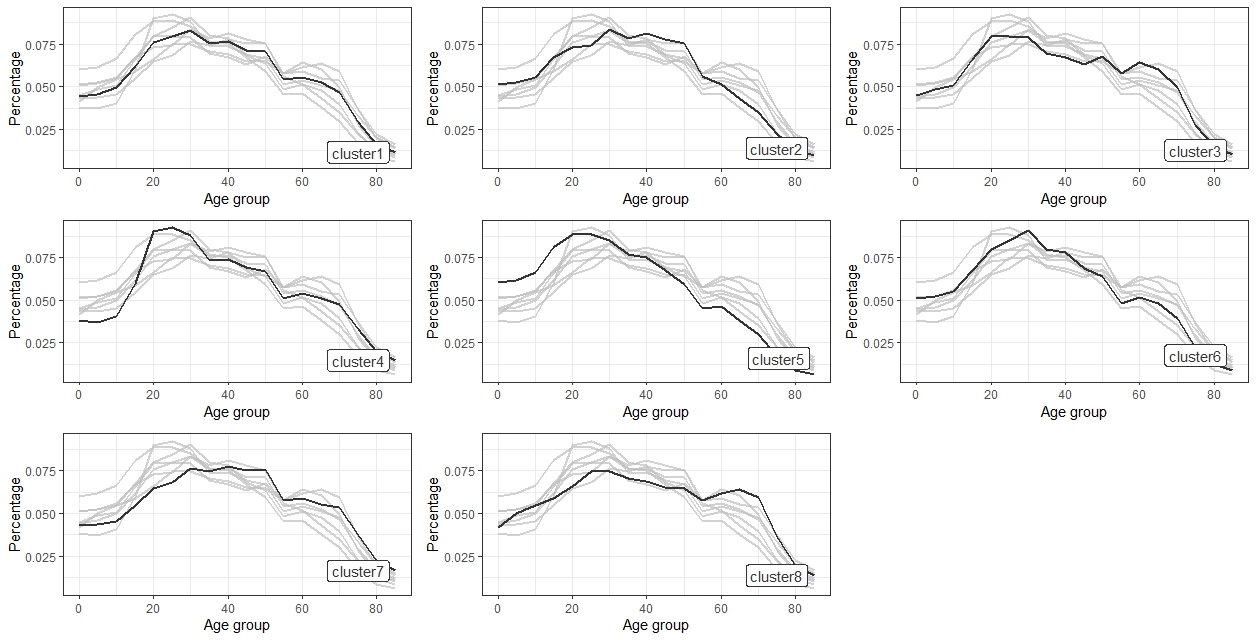}
\caption{Percentage of residents in each age group for each cluster \label{pub3}}
\end{figure}

\begin{figure}[ht]
\begin{center}
\includegraphics[scale=0.42]{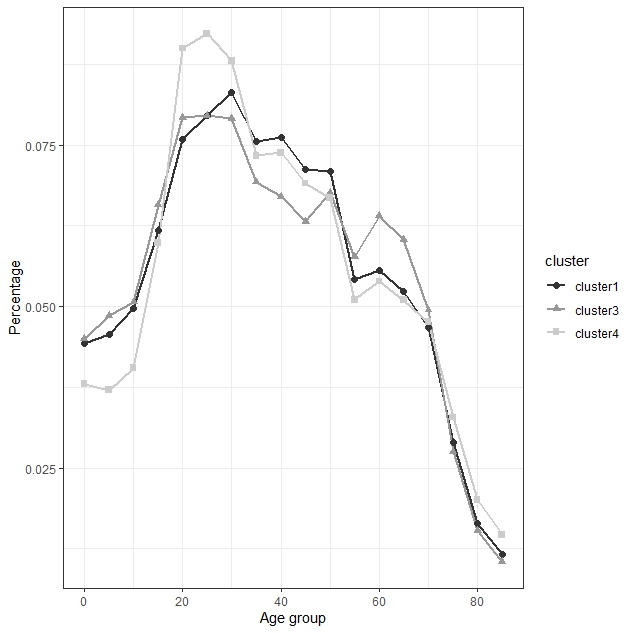}
\caption{Percentage of residents in each age group for three clusters \label{g1}}
\end{center}
\end{figure}

\begin{figure}[ht]
\begin{center}
\includegraphics[scale=0.42]{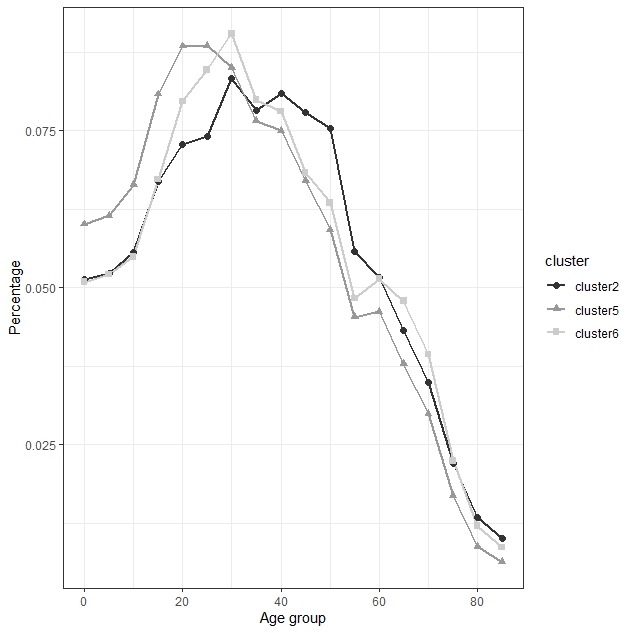}
\caption{Percentage of residents in each age group for three clusters \label{g2}}
\end{center}
\end{figure}

\begin{figure}[ht]
\begin{center}
\includegraphics[scale=0.42]{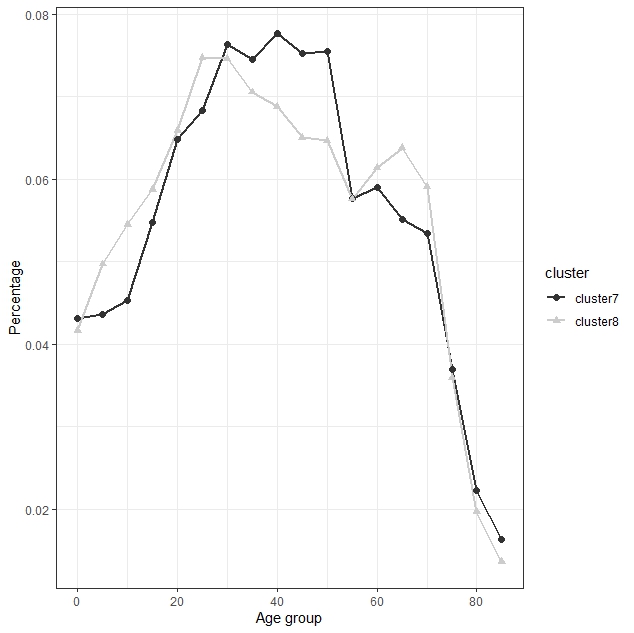}
\caption{Percentage of residents in each age group for two clusters \label{g3}}
\end{center}
\end{figure}

\begin{figure}[ht]
\begin{center}
\includegraphics[scale=0.42]{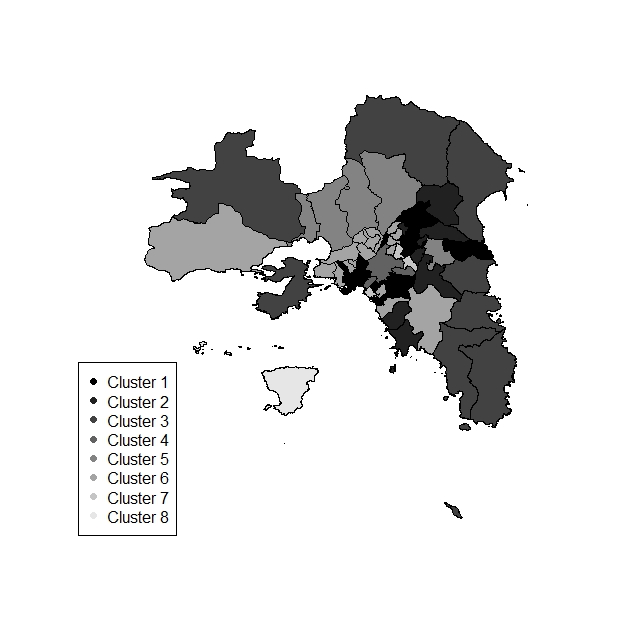}
\caption{Map of clusters according to eight component models \label{pub2}}
\end{center}
\end{figure}

The two models propose the same clusters that can be seen in Figure \ref{pub2}, while the parameters of each cluster are depicted in Figure \ref{pub3}. Table \ref{tab.6} includes the mean age of each group. Based on the mean age, clusters with similar value have been grouped in order to present the proportion of residents in each age group, for every cluster. The results are presented in Figures \ref{g1}, \ref{g2} and \ref{g3}. It seems that there is an interesting structure of municipalities of Attica with respect to age of the residents. \par

\begin{table}[ht]
  \centering 
  \begin{tabular}{cc}
  \hline
  Cluster & Mean age \\
 \midrule 
 1 & 37.57\\    
 2 & 35.87\\
 3 & 37.60\\
 4 & 38.22\\
 5 & 32.69\\
 6 & 35.40\\
 7 & 39.64\\
 8 & 39.04\\
  \bottomrule
 \end{tabular} 
  \caption{Mean age of residents in each cluster. \label{tab.6}}
\end{table}

 {\textbf{Clusters 1,3,4}:}
Clusters 1,3 and 4 represent municipalities that have an intermediate structure compared to the regions of the other clusters. Clusters 1 and 3 seems to have similar structure for the ages 0-25 when cluster 4 has lower proportions on ages 0-10. For ages 20-30, there is a peak for cluster 4. Looking at the map (\ref{pub2}), cluster 4 includes centre of Attica (Athens) with two other adjacent areas. In these municipalities, there are universities and as a consequence it is plausible to have significant number of young people, especially of these ages. Based on Figure \ref{pub3}, the peak in these ages is the highest among all clusters.  For the ages 30-70, cluster 1 and 4 have similar behaviour. According to the map, municipalities of cluster 4 are closer to municipalities of cluster 1. Given that cluster 1 is the central part of Attica and that the referred age groups imply the workforce of each region, the similarity is plausible, providing that work opportunities are better in the center than in the suburbs. As ages of 70+ are concerned, the differences are meaningful for the three clusters.       

 {\textbf{Clusters 2,5,6}:}
Clusters 2,5 and 6 consists of municipalities with the youngest population among municipalities of Atttica. Although the three clusters have similar population structure, the interesting is in cluster 5. As it is obvious from Figure \ref{g2}, cluster five has the highest proportions in ages 0-25 and the lowest values in ages 50+. In fact, the above annotations can be generalized as a comparison of cluster 5 with all the other clusters of the analysis (Figure \ref{pub3}). The essential difference of this cluster is attributed to the excess presence of Roma in these regions.

 {\textbf{Clusters 7,8}:}
Clusters 7 and 8 constitute a very small part of Attica since they consists of three and one municipalities respectively. These clusters presents slightly lower percentages in younger population and higher records on elderly, compared with the rest clusters. As a result, mean age of these clusters is somewhat higher from the rest. Cluster 7 is close to centre (cluster 4) and this explains the high percentage of residents in ages correspond to workforce. As cluster 8 is concerned it refers to Aegina island. The chances for educational or work improvement is difficult to an island, so the excess of elderly population can be explained as it is known destination fro retired people.      

\section{Conclusion}
\label{sec6}
In this paper, we proposed an extension of the work by \cite{Alfo2008} and \cite{Alfo2009} in the context of multinomial mixture. The aim is to find an appropriate way to cluster geographical regions according to demographic traits and especially the age. It seems that in this way population structure of an area can be described and be studied in a more efficient way. The proposed model not only does it account for the common heterogeneity, but also for spatial heterogeneity that exists due to the nature of the data. 

The spatial information is incorporated by prior probabilities. More specifically, allocation variables are modelled as MRF. Then, model estimation is based on a modification of CEM, where a step for approximating the field is preceded before common steps of CEM. Simulated field algorithm has been selected rather than mean field or mode field, since literature supports that it is better.

Although the usefulness of this model, there are issues to be concerned. The initiazlization process is extremely demanding because of the excess of parameters to be estimated and as a result the algorithm is possible to be trapped in local maximum. Moreover, in case of large records in each category of the multinomial distributions, as in the examined example, computational problems may arise due to very small probabilities.

Finally, the presented model can be enriched by the usage of more than one demographics. For example, it would be interesting using not only age but also sex, with the purpose of gaining more information about the population structure. In addition, prior probabilities can also be improved by inserting extra explanatory variables except for neighbors' behavior.

\bibliographystyle{chicago}
\bibliography{masterbiblio}

\end{document}